\documentclass[prd,amsmath,amssymb,floatfix,superscriptaddress,nofootinbib,twocolumn]{revtex4-1}

\def\be{\begin{equation}}
\def\ee{\end{equation}}
\def\bea{\begin{eqnarray}}
\def\eea{\end{eqnarray}}
\newcommand{\vs}{\nonumber\\}

\newcommand{\ec}[1]{Eq.~(\ref{eq:#1})}

\newcommand{\eql}[1]{\label{eq:#1}}
\newcommand{\sfig}[2]{
\includegraphics[width=#2]{#1}
        }

\newcommand{\Sfig}[2]{
   \begin{figure}[thbp]
   \begin{center}
    \sfig{#1.pdf}{\columnwidth}
    \caption{{\small #2}}
    \label{fig:#1}
     \end{center}
   \end{figure}
}

\newcommand{\rf}[1]{\ref{fig:#1}}

\usepackage[utf8]{inputenc}
\usepackage{graphicx}
\usepackage{amssymb}
\usepackage{amsmath}
\usepackage{bm}
\usepackage{color}
\usepackage{enumitem}
\usepackage[linktocpage=true]{hyperref} 
\hypersetup{
    colorlinks=true,       
    linkcolor=red,         
    citecolor=blue,        
    filecolor=magenta,      
    urlcolor=blue           
}
\usepackage[all]{hypcap} 

\definecolor{darkgreen}{cmyk}{0.85,0.1,1.00,0}

\begin{document}

\title{Distortions in the Surface of Last Scattering}

\author{\large Peikai Li}
\author{\large Scott Dodelson}
\affiliation{Department of Physics, Carnegie Mellon University, Pittsburgh, Pennsylvania 15312, USA}
\author{\large Wayne Hu}
\affiliation{Kavli Institute for Cosmological Physics, Department of Astronomy \& Astrophysics,
Enrico Fermi Institute, The University of Chicago, Chicago, IL 60637, USA}

\date{\today}

\begin{abstract}
The surface of last scattering of the photons in the cosmic microwave background is {\bf not} a spherical shell. Apart from its finite width, each photon experiences a different gravitational potential along its journey to us, leading to different travel times in different directions. Since all photons were released at the same cosmic time, the photons with longer travel times started farther away from us than those with shorter times. Thus, the surface of last scattering is corrugated, a deformed spherical shell. We present an estimator quadratic in the temperature and polarization fields that could provide a map of the time delays as a function of position on the sky. The signal to noise of this map could exceed unity for the dipole, thereby providing a rare insight into the universe on the largest observable scales.
\end{abstract}

\maketitle

\section{Distance to the Last Scattering Surface}
\newcommand\fd{d}

The theory of general relativity dictates that particles traveling through gravitational potential wells experience time delays~\cite{1964PhRvL..13..789S}. If two photons are emitted at the same time, then they will travel different distances depending upon the potential $\Phi$ through which they travel. In the cosmological context of an expanding,  spatially flat background, the 
fractional difference in comoving distance
$D_*$ to a source at redshift $z_*$ is 
\be
d(\hat n) =- \frac{2}{D_*}\, \int_0^{D_*} dD\, \Phi\left(D \hat n; t(D)\right)\eql{defd},
\ee
where 
$t(D)$ is the age of the universe when the photon is a comoving distance $D$ from us, and  we use the space-time metric convention 
\bea
ds^{2}= -(1-2\Phi)dt^{2}+a^{2}(1+2\Phi)d\vec{x}^{2}
\eea
with $a(t)$ the scale factor.
Note the sign in \ec{defd}: if photons pass through an over-dense region where $\Phi>0$, then they experience a time delay and therefore they arrive from a closer distance than the unperturbed last scattering surface\footnote{There is also a geometric time delay that is typically of the same size for a single lens but is much smaller here on the large scales of interest.}.

Photons that comprise the cosmic microwave background (CMB) experience these same time delays or advances~\cite{Hu:2001yq} where $z_*$ is the redshift corresponding to the last scattering surface. Since photons do not decouple instantaneously from the electron-proton plasma, the surface of last scattering is often said to have a finite width, and a more accurate expression for the fractional difference in distance traveled is
\be
\fd(\hat n) = 2\, \int_0^{\infty} dz \, e^{-\tau(z)}\, K_\fd(z)\Phi\left(D(z)\hat n; t(z)\right) ,
\eql{deffield}
\ee
where $H(z)$ is the Hubble expansion rate; $K_\fd(z) =  -(H(z)D_*)^{-1}$ and $D_* =  \int_0^\infty dz' e^{-\tau(z') }/H(z')$.  Here 
 $\tau$ is the optical depth, ignoring reionization, which becomes very large at times smaller than the epoch of  last scattering, $t_*$ or equivalently when $z>z_*$.

This directional-dependent change in the distance to last scattering implies that the last scattering surface is not a simple spherical shell. There are two other well-studied phenomena that undercut the notion that the photons in the CMB freely streamed to us from a infinitely thin last scattering sphere. First, since the mean free path at recombination was finite, the last scattering surface has a finite width, and this is accounted for in all computations of CMB anisotropies. Second, the photons in the CMB experience angular deflections as they traverse the inhomogeneous universe~\cite{Hu:2001tn,Lewis:2006fu} and this effect has been exploited by recent experiments~\cite{Smith:2007rg,Ade:2013tyw,Story:2014hni,Sherwin:2016tyf,Aghanim:2018oex} that make maps of the projected gravitational potential.

Although deflections and delays are two different phenomena, they share some similarities, especially in the case of the CMB. Both are determined by the integrated potential along the line of sight, although with slightly different kernels, as depicted in Figure~\rf{kernel}: the integrated potential $\phi$ that determines deflections has the same form as the right-hand side of \ec{deffield}  with
\be
K_\phi(z) = \frac{D_*-D(z)}{ D(z) D_* H(z)} .
\ee
The corresponding auto and cross power spectra are shown in Figure~\rf{Spectrum}.
It is clear that they are highly anticorrelated, so as a first approximation, we might view the maps of the lensing potential created for example in \citet{Aghanim:2018oex} as  maps of distance to the last scattering surface. Another similarity, one that has not yet been exploited, is that the quadratic estimator formalism  \cite{Hu:2001tn} can be applied to the delays as well, and this is what we will do in this paper.  We start though with the rather daunting facts that the RMS fractional distance differences are a factor of ten smaller than the RMS angular deviations 
 and their impact on CMB power spectra is even smaller \cite{Hu:2001yq}. Further, while the latter peaks at degree scales, the former peak on the largest scales where cosmic variance is
 higher.

\Sfig{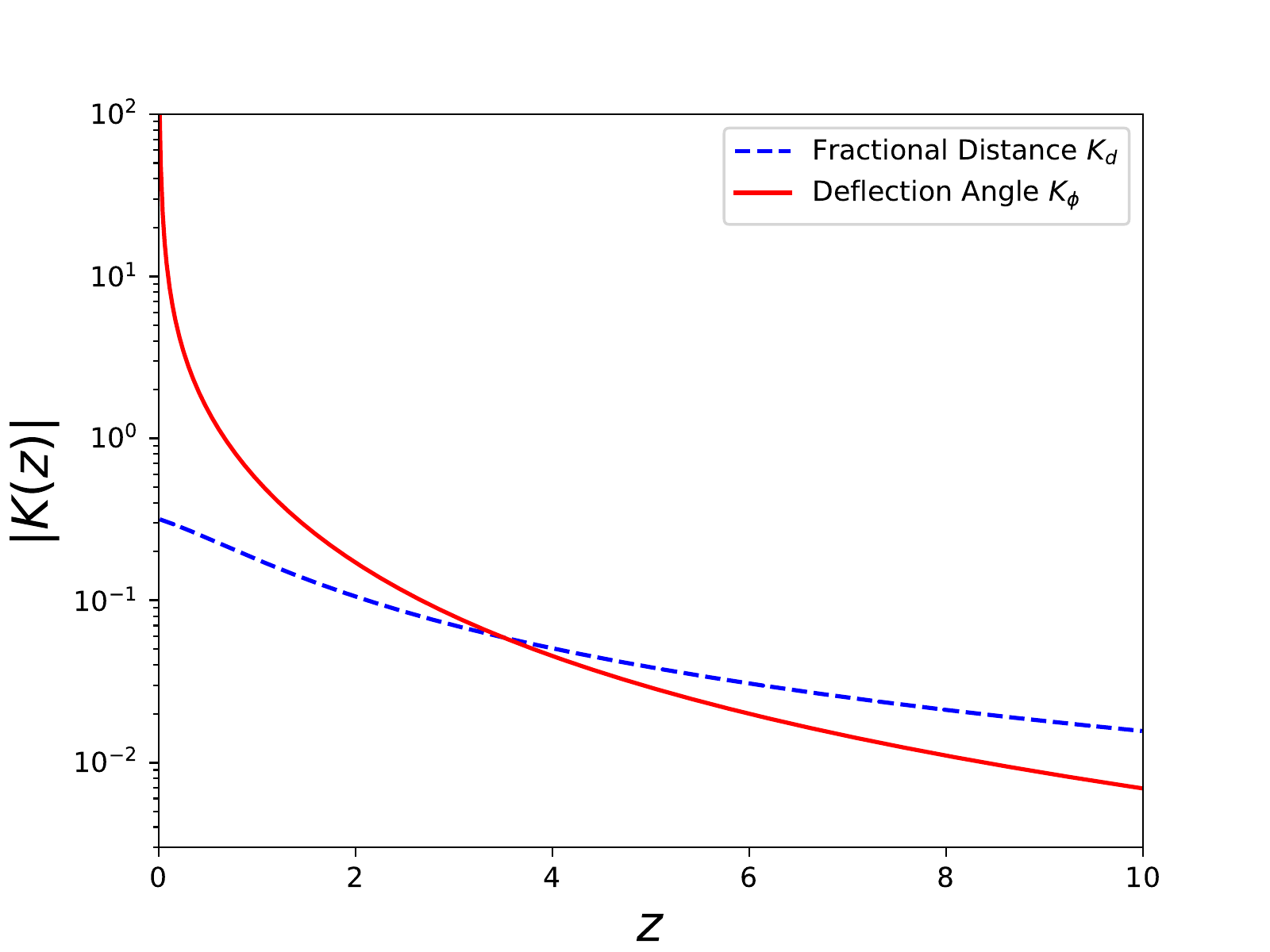}{Kernel that weights the integral of the gravitational potential for the time delay examined here and the more carefully studied deflection angle.}
\Sfig{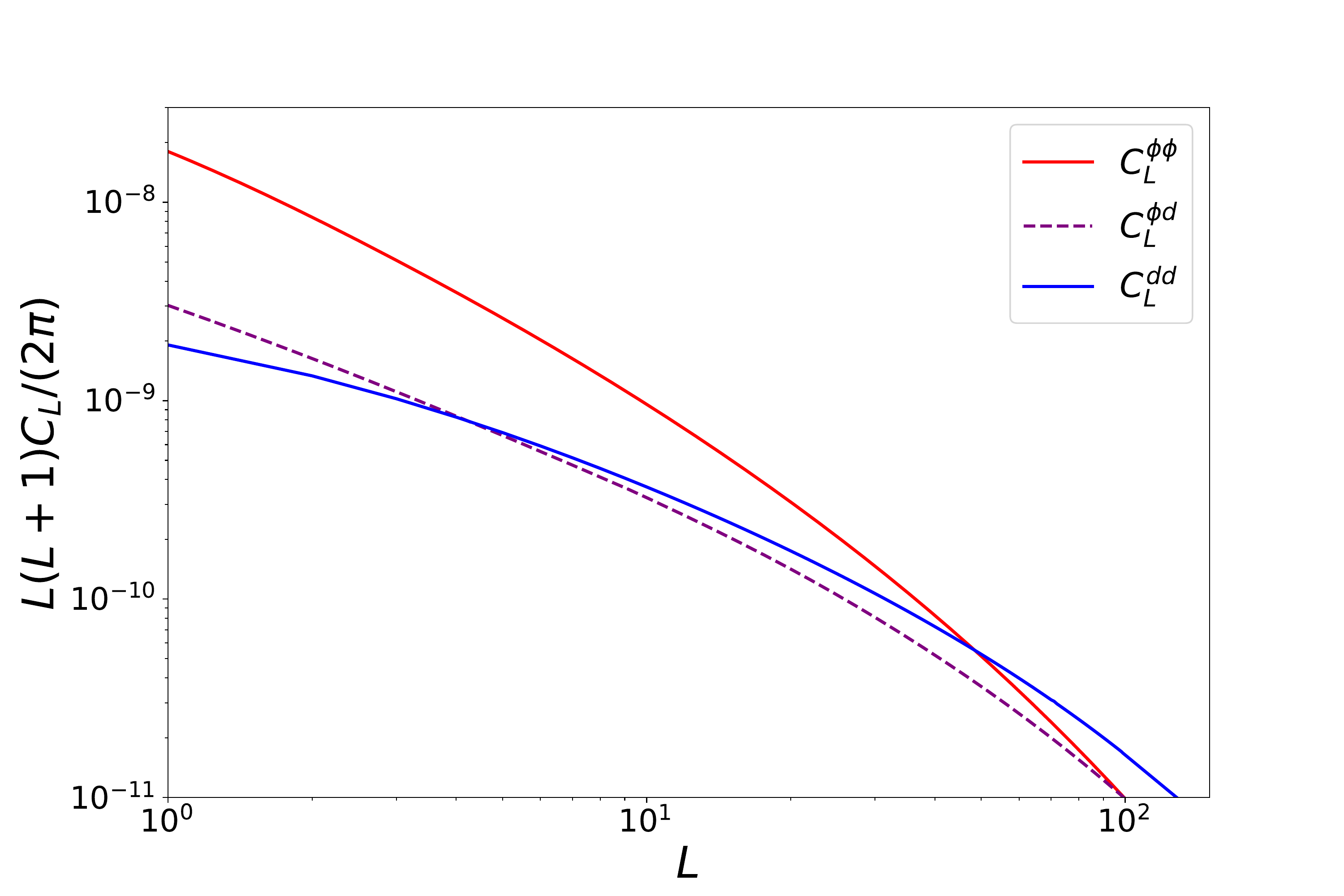}{Power spectra for the lensing deflection potential
($\phi\phi$), fractional distance ($dd$) and deflection-distortion cross correlation
($\phi d$). Dashed line means $C^{\phi d}_{L}$ is negative.}

\section{Effect of distance changes on the CMB}
\newcommand\dtl{\delta\Theta^{\rm defl}}
\newcommand\dtd{\delta\Theta^{\rm dist}}
\newcommand\tob{\Theta^{\rm obs}}
\newcommand\tu{\tilde{\Theta}}
\newcommand\td{\Theta^{\rm dist}}
The observed temperature in a given direction $\tob(\hat n)$ is the undistorted temperature $\tu(\hat n)$ plus the deflection due to gravitational lensing plus a term proportional to the small fractional difference $\fd(\hat n)$.  Linearizing these distortions, we obtain
\be
\tob(\hat n) = \tu(\hat n) + \nabla_{i} \phi(\hat n) \nabla^{i}\tu(\hat n) + \frac{\partial\tu(\hat n)}{\partial \ln D}\, \fd(\hat n),
\ee
where $D$ is the distance to the radiation sources, here mainly $D=D_*$ the distance 
to recombination.  We shall see below that we can express this radial derivative in
terms of operations on the radiation transfer function.
In harmonic space we can write
\be
\tob_{lm} = \tu_{lm} +\dtl_{lm}+ \dtd_{lm}
\ee
with the two first order terms due to deflection and the change in distance equal to
\bea
\dtl_{lm} &=& \sum_{LM}\sum_{l'm'}{}_{0}I_{lLl'}^{mMm'}\tu_{l'm'}\phi_{LM},\nonumber\\
\dtd_{lm} &=& \sum_{LM}\sum_{l'm'} {}_{0} J_{lLl'}^{mMm'}\frac{\partial \tu_{l'm'}}{\partial \ln D}d_{LM}\eql{dtd}
.\eea
Notice that both effects couple the undistorted temperature field to the observed
temperature field at a different multipole.
Here, we have written the integral over the product of three spherical harmonics as $_0I$ and $_0J$ to enable generalization to the case of polarization, which involves spin $s=2$ harmonics. The general expression is  
\bea
\eql{FG}
{}_{s}I_{lLl'}^{mMm'}  &=& (-1)^m\,
\bigl(\begin{smallmatrix} l & L & l' \\ -m & M & m'  \end{smallmatrix}\bigr) {}_{s}F_{lLl'}, \\
{}_{s}J_{lLl'}^{mMm'}  &=& (-1)^m\,
\bigl(\begin{smallmatrix} l & L & l' \\ -m & M & m'  \end{smallmatrix}\bigr) {}_{s}G_{lLl'}, \nonumber
\eea
with
\bea
{}_{s}F_{lLl'} &\equiv& \big[L(L+1)+l'(l'+1)-l(l+1)\big]\nonumber  \\ & \times&\sqrt{ \frac{ (2l+1)(2L+1)(2l'+1) }{16 \pi} }\begin{pmatrix} l & L & l'  \\   s & 0 & -s  \end{pmatrix} ,\nonumber\\
{}_{s}G_{lLl'} &\equiv& \sqrt{ \frac{(2l+1)(2L+1)(2l'+1) }{4 \pi}} \begin{pmatrix} l & L & l' \\  \vphantom{a^b} s & 0 & -s  \end{pmatrix}\eql{fg} 
.\eea
Note the extra two powers of the multipoles in the function $F$ that governs deflection; these follow from the fact that both the temperature and the potential are differentiated with respect to transverse position on the sky. By contrast, the radial derivative that governs the impact of the time delay, or change in distance to the last scattering surface, appears in \ec{dtd} as the logarithmic derivative of the undistorted coefficients $\tilde\Theta_{LM}$.

As in the case of the effect of deflections on the CMB, the varying distances to the last scattering surface leads to correlations between $l$-modes that differ from one another. 
First let us define the power spectrum of the undistorted fields
\begin{equation}
\tilde C_l^{\Theta\Theta} =  \frac{2}{\pi}\,\int_0^\infty k^2 dk P_{\cal R} T_l^ \Theta(k)T_l^ \Theta(k),
\eql{cdtt}
\end{equation}
where $T_l^\Theta(k)$ is the radiation transfer function and $P_{\cal R}$ is the  power spectrum of the initial comoving curvature field ${\cal R}$.  Note that the transfer function is a radial integral over
the sources at a distance $D$, projected onto multipole moment $l$.
We proceed as in Ref.~\cite{Hu:2001tn} by focusing on the expectation of off-diagonal (${l_1,m_1\ne l_2,m_2}$) terms quadratic in the observed moments:
\bea
\langle \tob_{l_1m_1} \tob_{l_2m_2} \rangle &=& \sum_{LM}  (-1)^M \bigl(\begin{smallmatrix} l_1 & l_2 & L \\ m_1 & m_2 & -M  \end{smallmatrix}\bigr) \nonumber \\
\vs&&\times\big[ \phi_{LM}f_{l_{1}Ll_{2}} + d_{LM}g_{l_{1}Ll_{2}} \big] ,  \eql{quad}
\eea
where
\bea
f_{l_1Ll_2} &\equiv& \big[ \tilde{C}_{l_{1}}^{\Theta\Theta} {}_{0}F_{l_{2}Ll_{1}} +\tilde{C}_{l_{2}}^{\Theta\Theta} {}_{0}F_{l_{1}Ll_{2}}\big],
\vs
g_{l_1Ll_2} &\equiv&\big[ \tilde{C}_{l_{1}}^{\Theta\Theta,d} {}_{0}G_{l_{2}Ll_{1}} +\tilde{C}_{l_{2}}^{\Theta \Theta,d} {}_{0}G_{l_{1}Ll_{2}}\big].\eql{lfg}
\eea
The change in distance to the last scattering  produces the spectrum
\be
\tilde{C}^{\Theta\Theta,d}_{l} \equiv \frac{2}{\pi}\,\int_0^\infty  k^2 dk P_{\cal R} T_l^\Theta(k) 
T_l^{\Theta,d}(k).
\eql{cld}
\ee
This expression is identical to \ec{cdtt} other than the replacement of one of the transfer functions with 
\begin{equation}
T_l^{\Theta,d} \equiv \frac{\partial T_l^\Theta}{\partial \ln D},
\end{equation}
where the derivative is taken inside of the integrals over the radiation sources by modifying
the public CAMB code. 
The two spectra are shown in Fig.~\rf{TT1}. 

To clarify the meaning of these terms, consider the large scale limit where the temperature
source is the Sachs-Wolfe effect on the recombination surface at $D_*$, 
 $\Theta = {\cal R}/5$.   Then
 \begin{equation}
 T_l^\Theta(k) = \frac{1}{5} j_l(k D_*), \quad  T_l^{\Theta,d} (k) = \frac{1}{5}
 \frac{\partial j_l(k D_*)}{\partial \ln D_*}.
\end{equation}
More generally the modification to CAMB involves replacing the appropriate Bessel function kernel of
the source projection with its log derivative \cite{Hu:2001yq}.

Note the difference between the two off diagonal correlations in \ec{quad}. Each involves a derivative. The one that governs deflections, $f$, involves a derivative with respect to the transverse directions so $F$ as defined in \ec{fg} has more powers of $l$ than does $G$. The function that governs changes in distances involved a radial derivative, and this shows up in the spectrum $\tilde{C}^{\Theta\Theta,d}_{l}$.

\Sfig{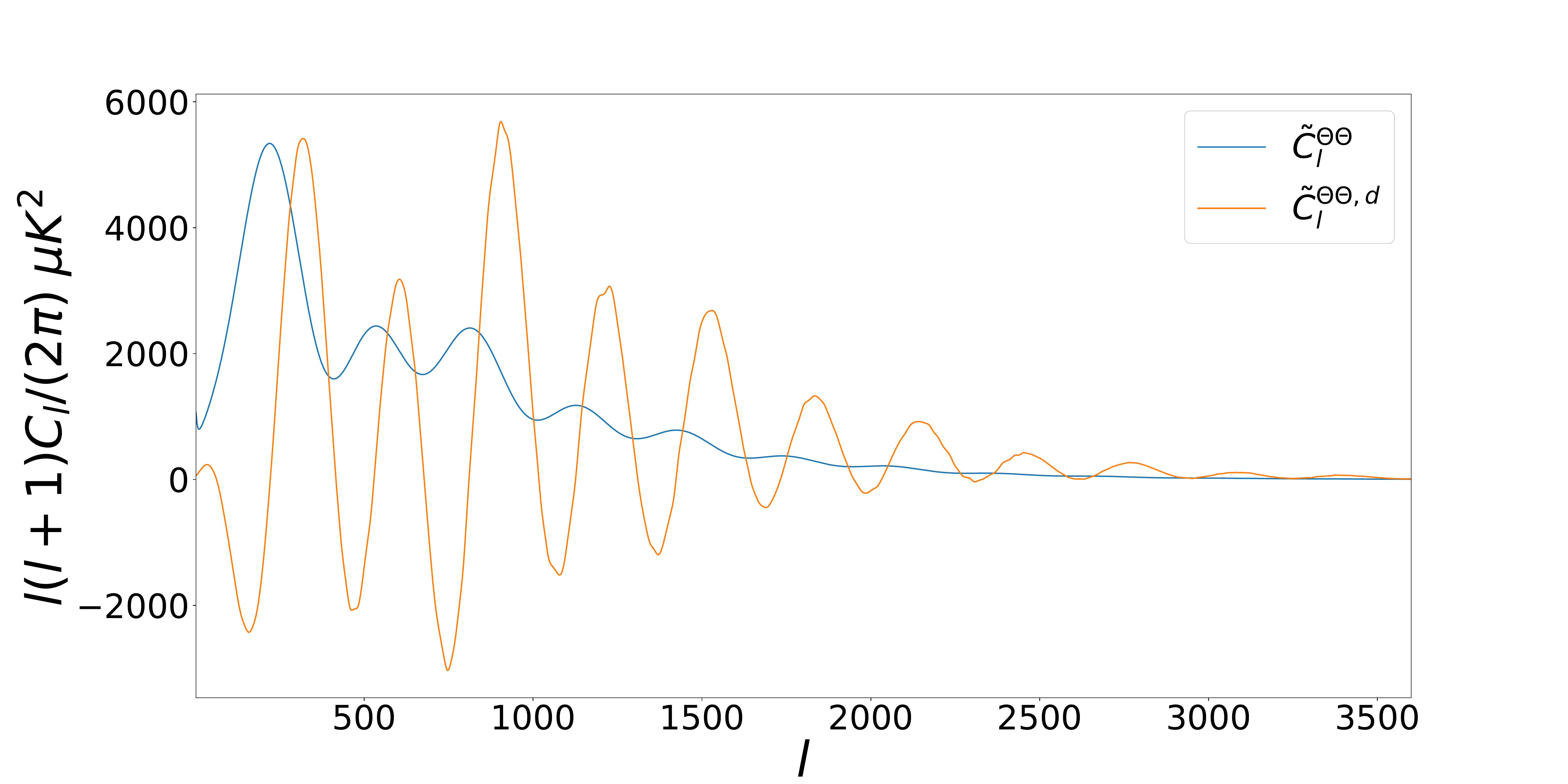}{Spectra of CMB temperature anisotropies and the logarithmic derivative of that spectrum with respect to the distance to the last scattering surface as defined in \ec{cld}
}

The correlation between different $l$-modes enables us, following Ref.~\cite{Okamoto:2003zw}, to extract information about the fields causing these correlations by forming quadratic estimators out of the observed temperature fields for both the gravitational potential responsible for deflections and the fractional distance field:
\bea
\hat \phi_{LM} &=& A_L 
\sum_{l_1m_1}\sum_{l_2m_2} 
 (-1)^M  \bigl(\begin{smallmatrix} l_1 & l_2 & L \\ m_1 & m_2 & -M  \end{smallmatrix}\bigr) 
 \nonumber \\  && \times 
 h^{\phi}_{l_1l_2}(L)  \tob_{l_1m_1} \tob_{l_2m_2}, \nonumber \\
\hat d_{LM} &=& B_{L} \sum_{l_1m_1}\sum_{l_2m_2}
(-1)^M  \bigl(\begin{smallmatrix} l_1 & l_2 & L \\ m_1 & m_2 & -M  \end{smallmatrix}\bigr) 
\nonumber \\ && \times
h^{d}_{l_1l_2}(L)  \tob_{l_1m_1} \tob_{l_2m_2}, \eql{distest}
\eea
where
\bea
h^{\phi}_{l_1l_2}(L)&\equiv& \frac{f_{l_1Ll_2}}{2C_{l_1}C_{l_2}} ,\nonumber\\
h^{d}_{l_1l_2}(L)&\equiv& \frac{g_{l_1Ll_2}}{2C_{l_1}C_{l_2}},\eql{defh}
\eea
and
\bea
A_L &\equiv& 
(2L+1)
 \left\{ \sum_{l_1l_2} h^{\phi}_{l_1l_2}(L)f_{l_1Ll_2}\right\}^{-1},\nonumber\\
B_L &\equiv& (2L+1) \left\{ \sum_{l_1l_2} h^{d}_{l_1l_2}(L)g_{l_1Ll_2}\right\}^{-1}.\eql{bt}
\eea 
With these definitions, 
$\langle \hat\phi_{LM} \rangle = \phi_{LM}$ when $d_{LM}=0$ and
$\langle \hat d _{LM} \rangle = d_{LM}$ when $\phi_{LM}=0$, where the average is over the undistorted CMB fields given fixed distortion fields.  To provide an optimistic bound
on detectability of the delay distortion, we ignore the cross contamination of the estimators for the time being. We return to this  issue in Sec.~\ref{sec:cross}.

The noise on these estimators is now given by the prefactors $A_L$ and $B_L$, so  
\be
\langle \hat d_{LM} \hat d^*_{L'M'}  \rangle = \delta_{LL'}\delta_{MM'} \left( C_L^{dd} + B_L \right)
\ee
with the first term on the right the signal and the second the noise. Fig.~\rf{Delay} shows the signal and noise at each $ L$ for several experimental configurations. Here, and throughout, the largest $l_{\rm max}$ we consider is 7000, as this seems to be within range being considered for a CMB-Stage 4 experiment (see Table 4.1 of Ref.~\cite{Abazajian:2016yjj}).

An estimate of the detectability of this signal can be obtained by computing the projected error, $\sigma_d$, on the amplitude $A^d$ of the power spectrum $A^dC_L^{dd}$, where the fiducial model has $A^d=1$.  Approximating the noise as Gaussian gives
\be
\left(\frac{1}{\sigma_d}\right)^2 = \sum_L^{\infty} \frac{(2L+1)f_{\rm sky}}{2} \left(\frac{C_L^{dd}}{C_L^{dd}+B_L}\right)^2,
\eql{sn}
\ee
where $f_{\rm sky}$ is the fraction of sky covered by the measurements.
Fig.~\rf{Delay} shows that most of the signal comes from the lowest $L$-modes, particularly $L=1$. However, even for a full-sky experiment and the most optimistic noise projections, the
auto power spectrum $C_L^{dd}$ will not be measurable using temperature only.

\Sfig{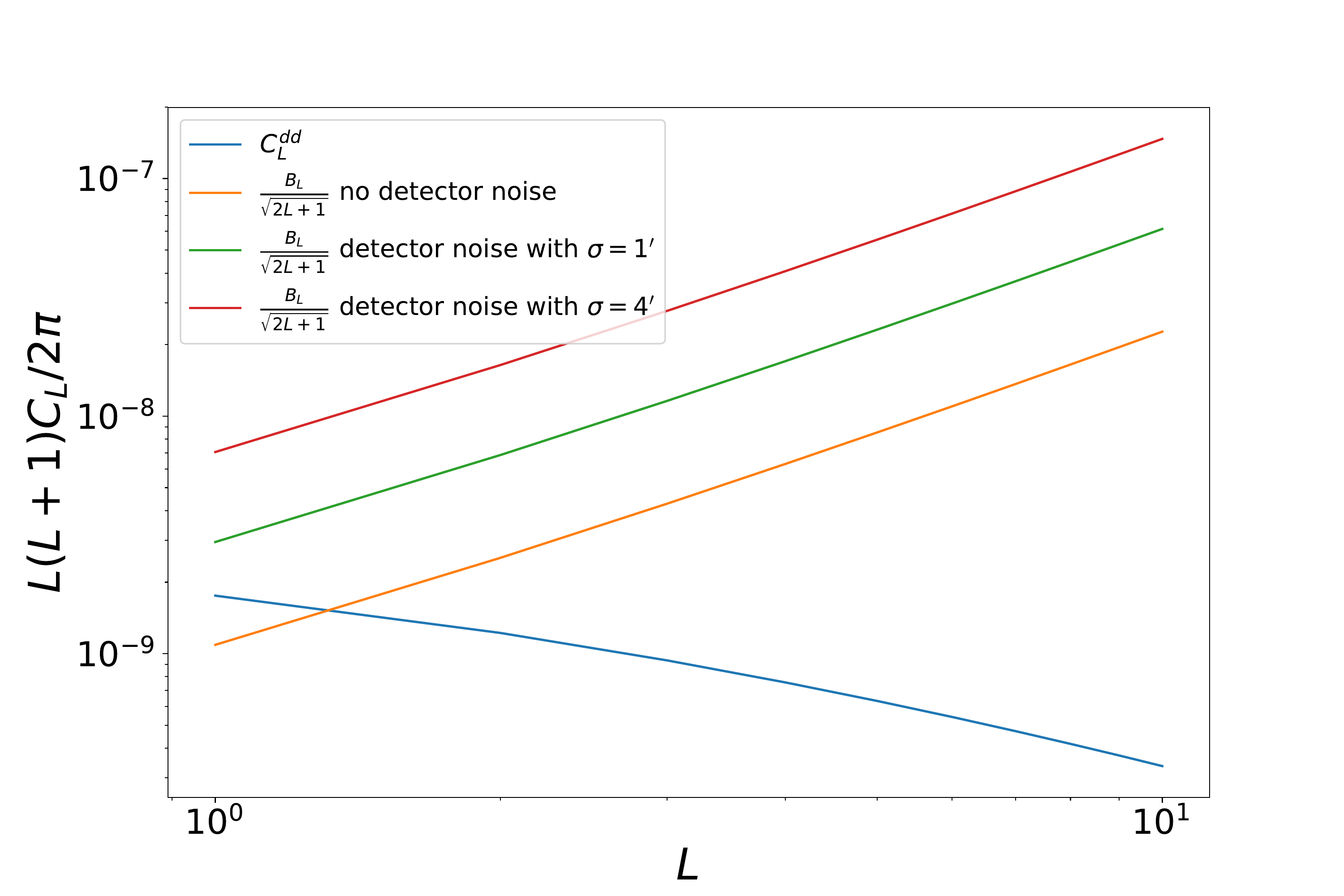}{Signal (decreasing blue curve) due to the distorted surface of last scattering and the noise using the quadratic estimator constructed from the small scale temperature anisotropy ($\Theta\Theta$ in the notation of Table 1) for several different experimental configurations. Most optimistic is no noise out to $l_{\rm max}=7000$; the other two noise curves have sensitivity of 1$\mu$K-arcmin and beam size. $\theta_{\rm FWHM}=1'$ or $4'$. Here, $f_{\rm sky}$ is set to one.}

\section{Polarization}

The estimator above used only the temperature anisotropy field, but the polarization field contains even more information about the lensing potential that governs deflection and distance changes. This was worked out in detail by Ref.~\cite{Okamoto:2003zw} for deflection, and we follow their notation here. There are now three fields of interest: temperature $\Theta$, and the two fields associated with polarization, $E$ and $B$. With letters $a,b$ each ranging over these three fields, we have 
\bea
\langle a^{\rm obs}_{l_{1}m_{1}}b^{\rm obs}_{l_{2}m_{2}}\rangle &=& \sum_{LM}(-1)^{M}\bigl(\begin{smallmatrix} l_1 & l_2 & L \\ m_1 & m_2 & -M  \end{smallmatrix}\bigr)\nonumber \\
&&\times
\big[ \phi_{LM}f^{\alpha}_{l_{1}Ll_{2}} + d_{LM}g^{\alpha}_{l_{1}Ll_{2}} \big] 
.\eea
The functions $f^\alpha$ and $g^\alpha$ are the generalizations of \ec{lfg} to include polarization (\ec{lfg} now corresponds to $\alpha=\Theta\Theta$). The full set of $f^{\alpha}$ was determined by Ref.~\cite{Okamoto:2003zw} and is reproduced in Table 1, which now includes the full set of $g^{\alpha}$ that govern the impact of changing radial distances. Note
that
\bea
\tilde{C}^{ab}_{l} \equiv \frac{2}{\pi}\,\int_0^\infty k^{2}dk \,&&{P}_{R}(k) T^{a}_l(k) T^b_l(k) \eea
denotes the power spectra of the undistorted fields with $T_l^a$ as the radiation transfer
function for the field $a$ and 
\be
\tilde{C}^{ab,d}_{l} \equiv  \frac{2}{\pi}\,\int_0^\infty k^{2}dk \,{P}_{R}(k)\frac{T_l^{a}(k)T^{b,d}(k)+T_l^{a,d}(k)T_l^{b}(k)}{2} 
\ee
with 
\begin{equation}
T_l^{a,d} \equiv \frac{\partial T_l^a}{\partial \ln D}
\end{equation}
again computed by modifying CAMB.  
 See Ref.~\cite{Hu:1997de} for a more detailed discussion.
Note that the angular deflection coefficients $f^\alpha$ do not carry superscript $^d$ because the derivatives are transverse and therefore captured by powers of $\ell$.

\begin{table}[thbp]
\scalebox{0.9}{
\begin{tabular}{|l|c|c|}
\hline
$\alpha$ & $f_{l_{1}Ll_{2}}^{\alpha}$ & $g_{l_{1}Ll_{2}}^{\alpha}$  \\
\hline
$\Theta \Theta$ & $\tilde{C}_{l_{1}}^{\Theta\Theta} {}_{0}F_{l_{2}Ll_{1}}+\tilde{C}_{l_{2}}^{\Theta\Theta}{}_{0}F_{l_{1}Ll_{2}}$ & $\tilde{C}_{l_{1}}^{\Theta\Theta,d} {}_{0}G_{l_{2}Ll_{1}}+\tilde{C}_{l_{2}}^{\Theta\Theta,d}{}_{0}G_{l_{1}Ll_{2}}$\\
$\Theta E$ &$\tilde{C}_{l_{1}}^{\Theta E} {}_{2}F_{l_{2}Ll_{1}}+\tilde{C}_{l_{2}}^{\Theta E}{}_{0}F_{l_{1}Ll_{2}}$ & $\tilde{C}_{l_{1}}^{\Theta E,d} {}_{2}G_{l_{2}Ll_{1}}+\tilde{C}_{l_{2}}^{\Theta E,d}{}_{0}G_{l_{1}Ll_{2}}$ \\
$EE$ &$\tilde{C}_{l_{1}}^{EE} {}_{2}F_{l_{2}Ll_{1}}+\tilde{C}_{l_{2}}^{EE}{}_{2}F_{l_{1}Ll_{2}}$ & $\tilde{C}_{l_{1}}^{EE,d} {}_{2}G_{l_{2}Ll_{1}}+\tilde{C}_{l_{2}}^{EE,d}{}_{2}G_{l_{1}Ll_{2}}$ \\
$\Theta B$ & $i\tilde{C}^{ \Theta E}_{l_{1}} {}_{2}F_{l_{2}Ll_{1}}$ &$i\tilde{C}^{\Theta E,d}_{l_{1}}{}_{2}G_{l_{2}Ll_{1}}  $\\
$EB$ &$i\big[\tilde{C}_{l_{1}}^{EE} {}_{2}F_{l_{2}Ll_{1}}-\tilde{C}_{l_{2}}^{BB}{}_{2}F_{l_{1}Ll_{2}}$\big] & $i\big[\tilde{C}_{l_{1}}^{EE,d} {}_{2}G_{l_{2}Ll_{1}}-\tilde{C}_{l_{2}}^{BB,d}{}_{2}G_{l_{1}Ll_{2}}\big]$ \\
$BB$ &$\tilde{C}_{l_{1}}^{BB} {}_{2}F_{l_{2}Ll_{1}}+\tilde{C}_{l_{2}}^{BB}{}_{2}F_{l_{1}Ll_{2}}$ & $\tilde{C}_{l_{1}}^{BB,d} {}_{2}G_{l_{2}Ll_{1}}+\tilde{C}_{l_{2}}^{BB,d}{}_{2}G_{l_{1}Ll_{2}}$ \\
\hline
\end{tabular}}
\caption{Explicit forms for $f$ anb $h$ of various polarizations. Notice that for $\Theta\Theta$, $\Theta E$, $EE$ and $BB$ polarization these functions are ``even"; for $\Theta B$ and $EB$ polarization they are ``odd" instead. ``Even" and ``Odd" indicate that the functions are non-zero only when $l_{1}+l_{2}+L$ are even or odd, respectively.}
\end{table}

An estimator can now be constructed for each of the pairs of fields, so letting $\alpha$ denote pairs of fields $(ab)$, we have
\bea
\hat{d}^{\alpha}_{LM} &=&\nonumber  (-1)^{M} B_{L}^{\alpha}\sum_{l_{1}m_{1}}\sum_{l_{2}m_{2}}\bigl(\begin{smallmatrix} l_1 & l_2 & L \\ m_1 & m_2 & -M  \end{smallmatrix}\bigr) \\
&& \times h^{\alpha,d}_{l_{1}l_{2}}(L) a^{\rm obs}_{l_{1}m_{1}}b^{\rm obs}_{l_{2}m_{2}} ,
\eea
where the minimum variance weights $h$ are generalizations of \ec{bt} 
\be
h^{\alpha=(ab),d}_{l_{1}l_{2}}(L) 
= \frac{C_{l_{2}}^{aa}C_{l_{1}}^{bb}g^{\alpha*}_{l_{1}Ll_{2}}-(-1)^{L+l_{1}+l_{2}}C_{l_{1}}^{ab}C_{l_{2}}^{ab}g^{\alpha*}_{l_{2}Ll_{1}}}{C_{l_{1}}^{aa}C_{l_{2}}^{aa}C_{l_{1}}^{bb}C_{l_{2}}^{bb}-(C_{l_{1}}^{ab}C_{l_{2}}^{ab})^{2}}.
\ee
Note that in the special cases $\alpha=aa$
\bea 
h_{l_{1}l_{2}}^{\alpha=(aa),d}(L)= \frac{g_{l_{1}Ll_{2}}^{\alpha*}}{2C_{l_{1}}^{aa}C_{l_{2}}^{aa}}. 
\eea 
and when $C_{l}^{ab}=0$ (e.g., for $\Theta B$ or $EB$),
\be
h^{\alpha,d}_{l_{1}l_{2}}(L) \rightarrow \frac{g^{\alpha *}_{l_{1}Ll_{2}}}{C_{l_{1}}^{aa}C_{l_{2}}^{bb}}. 
\ee
The covariance of these quadratic estimators
\be
\langle \hat{d}^{\alpha*}_{LM}d^{\beta}_{L'M'}\rangle \equiv \delta_{LL'}\delta_{MM'}\big[ C_{L}^{dd}+N_{L}^{d,\alpha \beta} \big]
\ee
with Gaussian noise given by
\bea
N_{L}^{d,\alpha\beta}&=&\frac{B_{L}^{\alpha*}B_{L}^{\beta}}{2L+1}\sum_{l_{1}l_{2}}  \left\{ h_{l_{1}l_{2}}^{\alpha,d*} (L)\big[ C_{l_{1}}^{ac}C_{l_{2}}^{bd}h_{l_{1}l_{2}}^{\alpha,d}(L)\right. \nonumber \\
&&\left. +(-1)^{L+l_{1}+l_{2}}C_{l_{1}}^{ad}C_{l_{2}}^{bc} h_{l_{2}l_{1}}^{\beta,d}(L)  \big]\right\} \eql{full} 
\eea
with $\alpha=(ab)$, $\beta=(cd)$. For $\alpha=\beta$, \ec{full} reduces to $N_{L}^{d,\alpha\alpha}=B_{L}^{\alpha}$.
Armed with these expressions, we can form a minimum variance estimator
\be
\hat{d}^{\rm mv}_{LM} = \sum_{\alpha}\omega^{d,\alpha}(L)\hat{d}^{\alpha}_{LM}  
\eql{dmvw}
\ee
with weights and variance given by
\bea
\omega^{d,\alpha}(L) & =&N^{d}_{L} \sum_{\beta}(N_{L}^{d,-1})^{\alpha \beta} \eql{weight} \nonumber\\
N^{d}_{L} &=& \frac{1}{\sum_{\alpha\beta} (N_{L}^{d,-1})^{\alpha \beta}}
\eea
where $(N_{L}^{d,-1})^{\alpha\beta}$ are the elements of the inverse of the delay noise matrix given by \ec{full}, with matrix indices given by quadratic combinations.  Here and below
we denote the noise of the minimum variance combination with no indices for simplicity.  Analogous expressions with the superscript $^\phi$ apply for the lens potential estimators.

We saw in Fig.~\rf{Delay} that small scale temperature maps only are not sufficient to detect this signal convincingly. To assess the added information contained in the polarization field, we show the detectability in the form of $\sigma_d$ for the lowest $L$-modes ($L\leqslant 5$, which contributes essentially all of the signal) as a function $l_{\rm max}$ for a noiseless experiment in Fig.~\rf{StoN}.

\Sfig{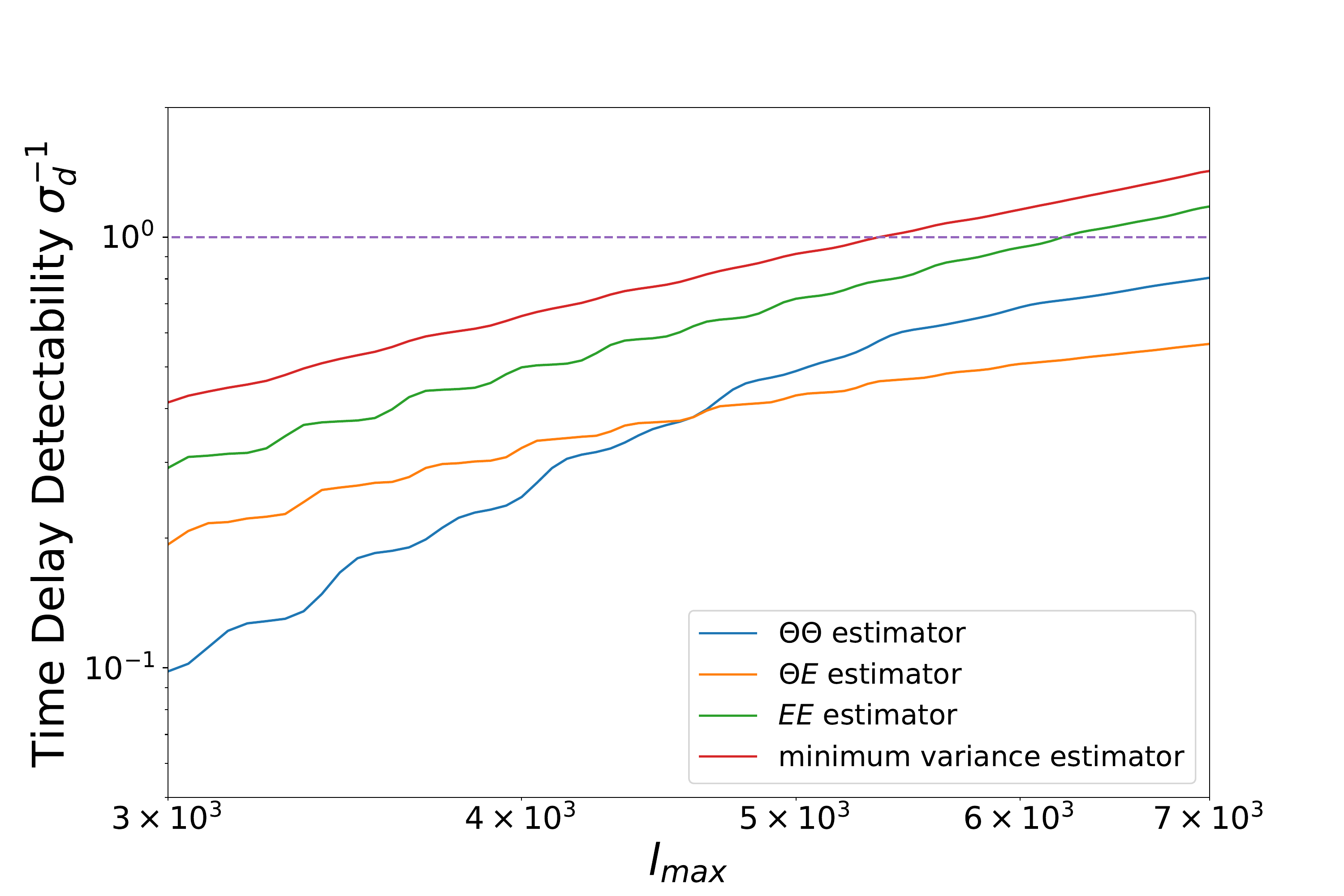}{Detectability of the spectrum of the time delay signal, $\sigma_d^{-1}$, using \ec{sn} for different quadratic estimators as a function of the largest mode, $l_{\rm max}$ accessed. Here we set $f_{\rm sky}$ to be $1$.}

Here we keep $l_{\rm min}$ fixed at $1000$ for the CMB fields (we tested that our final results are insensitive to this choice) and let $l_{\rm max}$ vary.
We can see that at $l_{\rm max} \approx 5000$, the minimum variance estimator detectability reaches 1, and at 7000 is 1.42 but of course this is for the most optimistic of configurations. 
Note that unlike the minimum variance lensing  estimator, the deflection estimator
gets little contribution from $EB$.   This is because the conversion of $E$ modes to $B$
modes is inefficient in the squeezed limit where $L\ll l_1,l_2$, as reflected in the difference between
$L+l_1+l_2$ even and odd configurations of the Wigner 3$j$ symbols, and that the $B$ modes from lensing
provide an intrinsic floor to detectability even in the absence of undistorted $B$ modes
from gravitational waves.

\section{Cross Power Spectrum}
\label{sec:cross}

The auto-spectrum of the distortion field, $d$, will apparently be very challenging to extract. Another possibility is to cross-correlate the quadratic estimator for the distortion field with other fields that are well-measured. Cross-correlations can be more easily detected if the two-fields are highly correlated and one of the fields can be detected with high signal to noise. Note from Figure 2 that $C^{\phi d}_{L}$ is negative and comparable to the auto spectra and so the two fields $\phi$ and $d$  are highly anticorrelated.

As a first attempt, we consider the cross correlation of the $d$-field with the $\phi$-field responsible for deflections. Without further optimization, the cross-spectrum for the $\alpha=ab$ and
$\beta=cd$ quadratic estimators is 
\bea
 \langle \hat{\phi}^{\alpha*}_{LM}\hat{d}^{\beta}_{LM} \rangle= {C}_{L}^{\phi d} +N_{L}^{c,\alpha\beta},\eql{crossn}
\eea 
where the superscript $c$ stands for cross. The noise in the estimators is also correlated because 
both estimators come from the  quadratic combinations of same observables 
\bea
N_{L}^{c,\alpha\beta}&=&\frac{A_{L}^{\alpha*}B_{L}^{\beta}}{(2L+1)}\sum_{l_{1}l_{2}}  \left\{ h_{l_{1}l_{2}}^{\alpha,\phi*} (L)\big[ C_{l_{1}}^{ac}C_{l_{2}}^{bd}h_{l_{1}l_{2}}^{\beta,d}(L)\right. \nonumber \\
&&\left. +(-1)^{L+l_{1}+l_{2}}C_{l_{1}}^{ad}C_{l_{2}}^{bc} h_{l_{2}l_{1}}^{\beta,d}(L)  \big]\right\}.
\eea
We can also construct the analogous noise cross spectrum for the separate minimum variance weighted estimators for $\phi$ and $d$  using
the weights of \ec{weight} 
\be
N^{\rm c}_L = \sum_{\alpha\beta } \omega^{d,\alpha}(L) \omega^{\phi,\beta}(L) N^{c,\alpha\beta}_L.
\ee

Assuming the noise is Gaussian, we can  estimate the fractional
error on the measurement of the amplitude of the cross spectra $C_l^{\phi d}$ from these minimum variance estimators as
\bea
\eql{sn1}
\frac{1}{\sigma_{\rm cross}^{2} }&=& \sum_{L=1}^{L_{\rm max}}(2L+1)f_{\rm sky} \\
&&\times\frac{(C_L^{\phi d})^2}{( C_L^{\phi d}+N_{L}^{ c})^2+
(C_L^{\phi\phi}+N_L^{\phi})(C_L^{dd}+N_L^{d})}. \nonumber
\eea
We show this quantity in Fig.~\rf{CrossStoN} where we have set
 $f_{\rm sky}=1$. We  see that in this ideal case, $S/N$ could reach about $2.5$. There are many other lens, and more generally large-scale structure tracers, that delay 
 reconstruction can be correlated with.  However the noise here is dominated by $N_L^d$, which is common to all such correlations, not
 $N_L^\phi$ or $N_L^c$.

 \Sfig{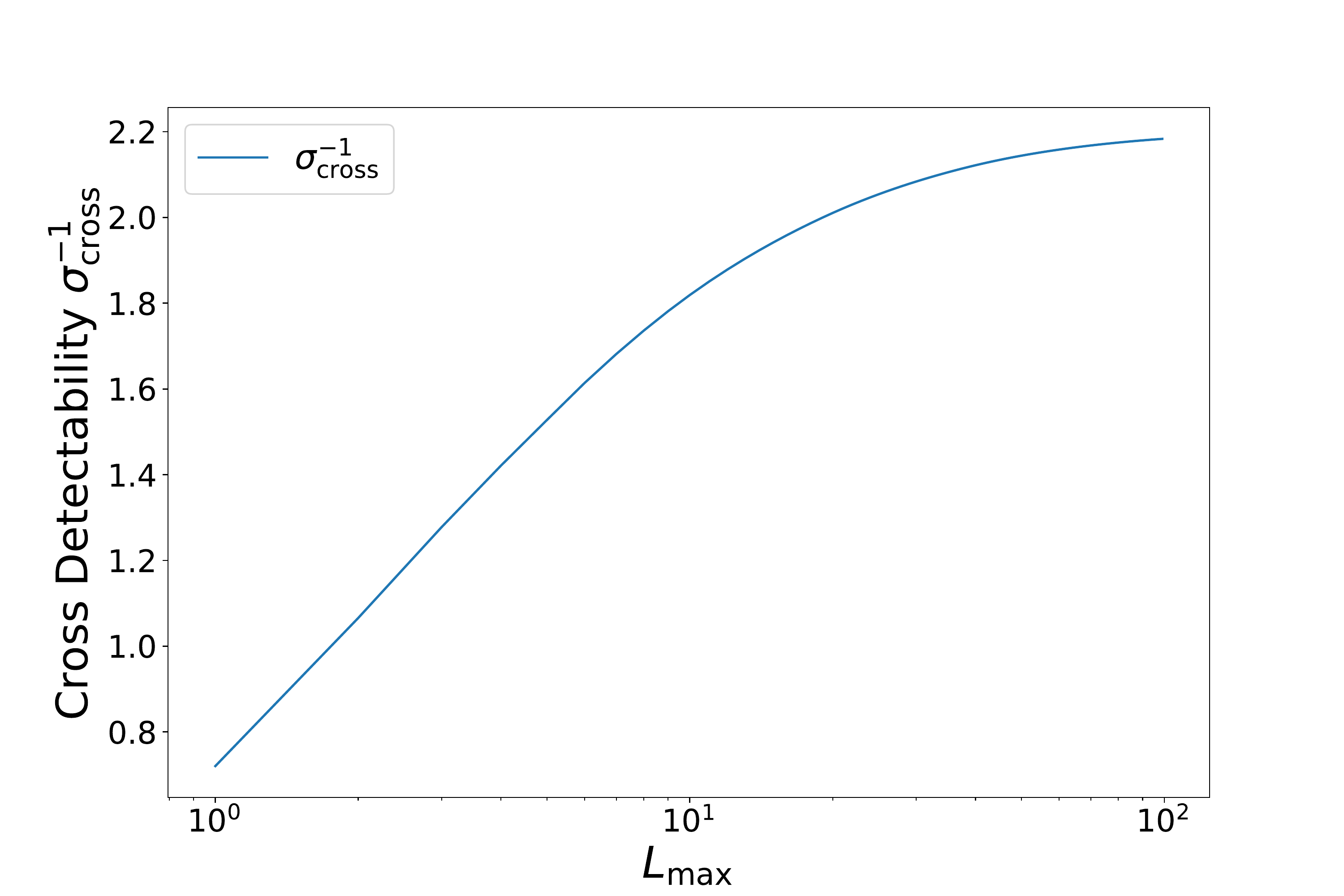}{The detectability of the cross power spectrum of the distance distortion field and the potential inducing deflections as a function of $L_{\rm max}$ (with  $l_{\rm max}=7000$). Unlike the auto-spectrum, there is signal out to $L_{\rm max}\sim 100$, but the contributions to the detectability plateau after that, so the best that can be hoped for with this cross spectrum is a $2.2\sigma$ detection.}

Up until this point, we have assumed that the estimators of $\phi$ and $d$ are not 
 contaminated by the other so as to provide the most optimistic bound on detectability of the
 delay distortion.   Since the cross power spectrum is potentially measurable at low $S/N$ we
 conclude by estimating this contamination.

 We can define the cross contamination by generalizing the computation in \ec{bt} to retain the distortion induced by $\phi$ on the estimator of $d$ and vice
 versa from \ec{quad} for the average over CMB modes
 \bea
 \langle \hat{d}^{\alpha}_{LM} \rangle &\equiv& d_{LM} +\phi_{LM}E_{L}^{\alpha}\sqrt{\frac{C_L^{dd}}{C_L^{\phi\phi}}}, \nonumber\\
 \langle \hat{\phi}^{\alpha}_{LM} \rangle &\equiv& \phi_{LM} +d_{LM}F_{L}^{\alpha}\sqrt{\frac{C_L^{\phi\phi}}{C_L^{dd}}},
\eea
where $E$ and $F$ measure the expected fractional contamination to each estimator in units of their uncontaminated signal rms
\bea
E^{\alpha}_{L}  \sqrt{\frac{C_{L}^{dd}}{C_{L}^{\phi\phi}}}  &=&\frac{\sum_{l_{1}l_{2}}h_{l_{1}l_{2}}^{\alpha,d}(L)f_{l_{1}Ll_{2}}^{\alpha}}{\sum_{l_{1}l_{2}}h_{l_{1}l_{2}}^{\alpha,d}(L)g_{l_{1}Ll_{2}}^{\alpha}} =\frac{N_{L}^{c,\alpha\alpha}}{N_{L}^{\phi,\alpha\alpha}},\nonumber\\
F^{\alpha}_{L} \sqrt{\frac{C_{L}^{\phi\phi}}{C_{L}^{dd}}}  &=&\frac{\sum_{l_{1}l_{2}}h_{l_{1}l_{2}}^{\alpha,\phi}(L)g_{l_{1}Ll_{2}}^{\alpha}}{\sum_{l_{1}l_{2}}h_{l_{1}l_{2}}^{\alpha,\phi}(L)f_{l_{1}Ll_{2}}^{\alpha}} =\frac{N_{L}^{c,\alpha\alpha}}{N_{L}^{d,\alpha\alpha}}.
\eea

We can then compute minimum variance estimator contaminations using \ec{dmvw} as
\bea
E_{L} &=&  
\sum_{\alpha}\omega^{d,\alpha}(L)  E^{\alpha}_{L} , \nonumber\\
F_{L} &=&\sum_{\alpha}\omega^{\phi,\alpha}(L) F^{\alpha}_{L} .
\eea

 In Fig.~\rf{Cont_multi} we show the relative contamination fields $E$ and $F$ for the minimum 
 variance estimators of $d$ and $\phi$ respectively.   The relative contamination $E$ for
 $d$ is large and increases with $L$, reflecting the similar structures in the
 mode coupling  of \ec{FG} but with extra factors of $L$ for $\phi$. 
 Conversely, the relative contamination $F$ for $\phi$ decreases with $L$: $F_{L=1} \approx -0.24$,  $F_{L=2} \approx -0.05$.  In principle lensing estimators 
 should be corrected for this effect at the lowest $L$. 
   While in the regime relevant to current 
 measurements, the delay contribution to the lens reconstruction is entirely negligible.

 If these cross contamination contributions are not removed at the reconstruction level, then
 the auto and cross spectrum measurements become biased since
 \bea
  \langle \hat{\phi}^{\rm mv*}_{LM}\hat{\phi}^{\rm mv}_{LM} \rangle &=&
  \left[ 1 +\frac{(F_L^{*}+F_L^{} )C_L^{\phi d}}{\sqrt{C_L^{\phi\phi} C_L^{dd} }}
  + F_L^{*}F_L^{} \right] C_L^{\phi\phi} 
  + N_L^{\phi}, \nonumber\\
    \langle \hat{d}^{\rm mv*}_{LM}\hat{d}^{\rm mv}_{LM} \rangle &=&
  \left[ 1 +\frac{(E_L^{*}+E_L^{} )C_L^{\phi d}}{\sqrt{C_L^{\phi\phi} C_L^{dd} }}
  + E_L^{*}E_L^{} \right] C_L^{dd} 
  + N_L^{d}, \nonumber\\
 \langle \hat{\phi}^{\rm mv*}_{LM}\hat{d}^{\rm mv}_{LM} \rangle&=& (1+ F_L^{*} E_L^{}){C}_{L}^{\phi d}+ \sqrt{C_{L}^{\phi\phi} C_{L}^{dd}} ( F_L^{*} + E_L^{}) \nonumber\\
&&  +N_{L}^{c} .
\eea
The large contributions from $E$ dominate the contamination of the delay auto and cross spectra. For the $F$ contamination to $C_l^{\phi\phi}$ measurements, even at the $L=1$, the renormalization factor in the brackets is about 1.3 and converges to unity rapidly with $L$. To the extent that noise and cross contamination can be neglected in the $\phi$ lensing measurements, the $E$ bias can in principle be removed at the $d$ reconstruction level without increasing sample variance for the power spectra.    In practice this would involve iterating
the estimators.   
However given that the mode 
coupling forms  of \ec{FG} are similar but not identical, especially in the different structures for $C_l^{ab}$ and $C_l^{ab,d}$, reweighting the $(L,l_1,l_2)$ triangles
to reduce the contamination would be more optimal.  Given the relatively low $S/N$ for 
even perfect removal, we leave these studies to a future work.

\Sfig{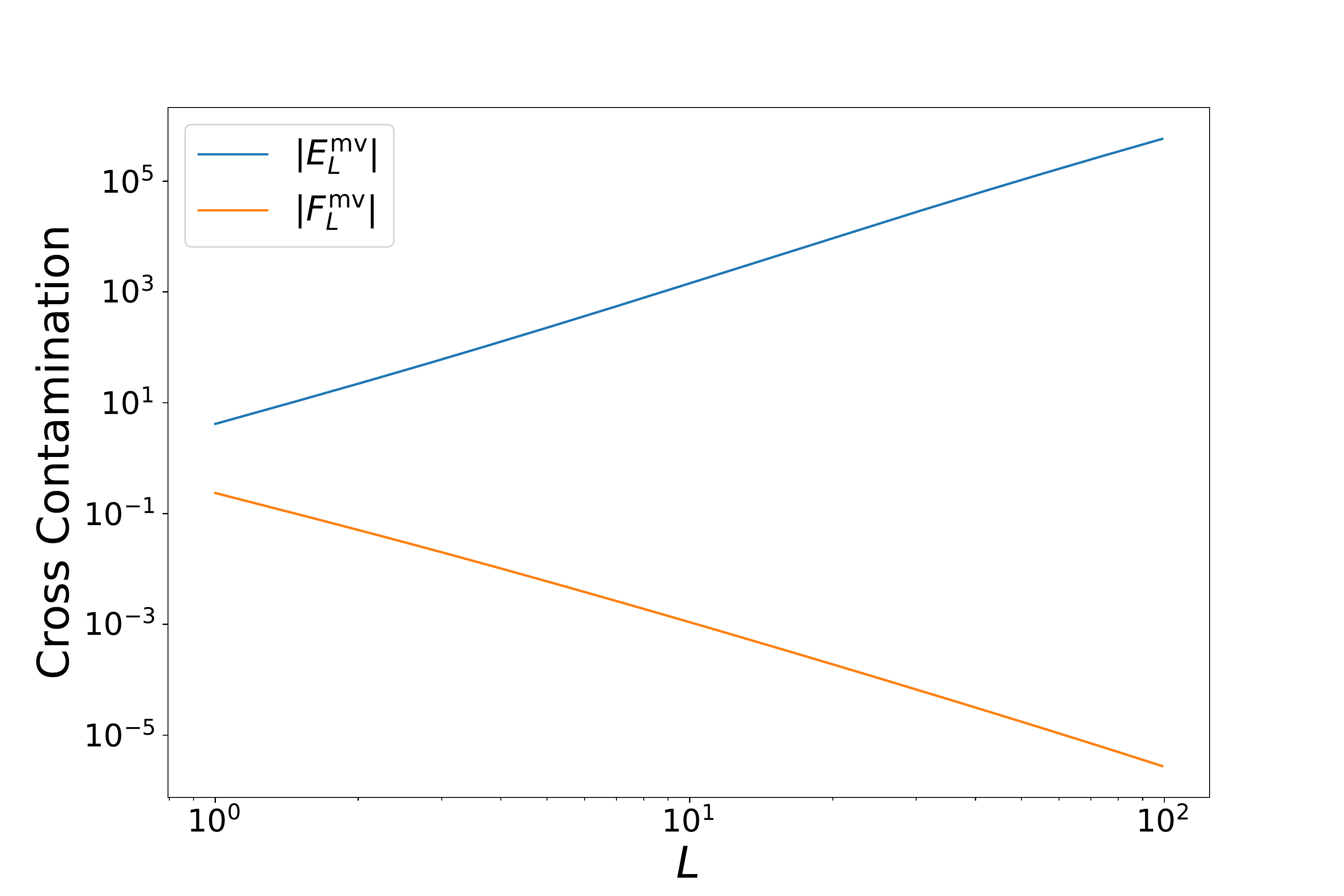}{Cross contamination $E$ for the fractional contribution of $\phi$ to the $d$ estimator and $F$ for the converse. $E$ and $F$ are both negative.}

\section{Conclusions}

The last scattering surface of the CMB is not purely spherical due to the different travel times experienced by photons as they traverse the inhomogeneous gravitational potential. In principle, these distortions in the distance to different directions is detectable, but we conclude here that the standard auto-correlation techniques will not be sufficient to enable detection in the near future. There is the possibility of cross-correlating a map of the distance distortions constructed with the quadratic estimators introduced here with another map of a closely related integrated potential and extracting the signal in that way. Indeed, this was the way that the transverse distortions in the CMB were first detected~\cite{Smith:2007rg}. Here, we have considered  the cross-correlation signal between the distance distortion and the standard transverse deviation maps and concluded that even an all-sky experiment with superior angular resolution would be detect the combination of auto and cross-spectra at less than 3-sigma. We leave exploration of other cross-correlations and the removal of cross contamination between lensing and delay estimators to future work.

To conclude, we emphasize that a measurement of the matter density on the largest observable scales, which is what a detection of the distance-distortion spectrum (either in auto- or cross-) would provide, carries the potential for enormous insight into the standard cosmological model. The apparent acceleration of the universe is of course a very large scale effect that remains a mystery. Inflation too is deeply embedded in the standard cosmological model and clues to it -- or its replacement -- might be found by studying the universe on the largest of scales. Over the past decades, a number of large scale anomalies have emerged (see, e.g., the first section of \citet{Hansen:2018pgg} for a review of the anomalies) and many ideas for models that might be responsible have emerged. A measurement of CMB deflections and time delays on the largest observable scales could help us either identify one such model or cast further doubt on the standard cosmological model.

Beyond the statistical limitations described in this paper, there are two caveats to the excitement of hunting for large scale physics by studying the distance-distortion. The first is the trivial comment that the deflection spectrum itself carries larger signal to noise even on the largest scales (although not by much, and of course two different measurements would be extremely worthwhile). More important is a physics question regarding the $L=1$ mode, the mode that carries the most signal to noise in the distance-distortion spectrum. In different contexts, there have been arguments that the dipole is suppressed by other effects~\cite{Zibin:2008fe,Itoh:2009vc}. A simple understanding of a large scale gradient in the gravitational potential (i.e., a dipole) would be that all matter experiences the same force and therefore velocity. A simple thought experiment of this ``bulk motion'' universe suggests that there would still be time delays of the sort described in this paper and the deflections that have been studied previously. However, this may be neglecting other effects that lead to cancellations. This issue too we leave for further study.

\acknowledgements

We thank Douglas Scott for useful discussions.  SD and PL were supported by U.S.\ Dept.\ of Energy contract DE-SC0019248.
 WH was supported by NASA ATP NNX15AK22G, U.S.\ Dept.\ of Energy contract DE-FG02-13ER41958
and  the Simons Foundation.


%\bibliography{refs}
\end{document}